\documentclass[aps,superscriptaddress,pra,reprint,longbibliography]{revtex4-2}
\usepackage{amsmath,amssymb,braket}
\usepackage{graphicx}
\usepackage{bm}
\usepackage{physics}
\usepackage{xcolor}
\usepackage{float}
\usepackage{booktabs}
\usepackage{tikz}
\usetikzlibrary{arrows.meta,positioning,decorations.pathreplacing}
\usepackage[colorlinks=true, allcolors=blue]{hyperref}

\usepackage[caption=false]{subfig}

\begin{document}

\title{{Dissipation-protected energy storage in interacting multilevel quantum batteries}}

\author{Yiğit Perçin}
\email{ypercin24@ku.edu.tr}
\affiliation{Department of Physics, Koç University, 34450 Sarıyer, Istanbul, Türkiye}

\author{Özgür E. Müstecaplıoğlu}
\email{omustecap@ku.edu.tr}
\affiliation{Department of Physics, Koç University, 34450 Sarıyer, Istanbul, Türkiye}
\affiliation{TÜBİTAK Research Institute for Fundamental Sciences (TBAE), 41470 Gebze, Türkiye}

\date{\today}

\begin{abstract}
Dark and subradiant states have emerged as a promising resource for stabilizing open quantum batteries against dissipation, but existing studies are largely limited to qubit ensembles and symmetry-based constructions. Here, we introduce a systematic, thermodynamically consistent framework for identifying long-lived energy storage states in interacting multilevel quantum batteries, combining the Davies master equation with a Morris--Shore (MS)–type decomposition of dissipative coupling blocks. Focusing on a minimal model of two interacting qutrits coupled to a common bath,
we analytically construct dark, bright, and funnel states—excited states that decay
exclusively into protected manifolds. We also derive quantitative robustness
conditions governed by the ratio of interaction strength to anharmonicity. We show
that multilevel ladder structure and exchange interactions enable energetic storage
states beyond the qubit case. Numerical simulations confirm that these states exhibit long-lived energy storage under typical Markovian dissipation. Finally, we show that high-energy funnel states provide a natural design target
for multilevel quantum batteries, as their decay pathways are highly structured
and directed toward protected manifolds. Knowledge and understanding of these pathways offer a
principled basis for developing future protection and control strategies in
superconducting multilevel platforms.

\end{abstract}

\maketitle

\section{Introduction}

Quantum batteries—quantum systems designed to store and deliver energy—have attracted increasing attention as potential applications of quantum coherence, correlations, and collective effects~\cite{Hovhannisyan2013, Alicki2013, FerraroNature2026}. Early theoretical studies demonstrated that collective charging protocols can enhance charging power relative to classical bounds~\cite{Campaioli2024, Binder2015, Campaioli2017, Ferraro2018}, while subsequent work emphasized the critical role of dissipation in determining practical battery performance.

In realistic settings, quantum batteries operate as open systems and inevitably suffer from energy leakage and self-discharge~\cite{Barra2019, Hovhannisyan2020}. A particularly promising strategy to mitigate these effects is the use of dark or subradiant states~\cite{Tejero2021, Quach2020, Liu2019}, which are protected against dominant decay channels by destructive interference or symmetry. Such states have been proposed and employed as long-lived storage manifolds in qubit-based batteries, excitonic systems, and open-system charging protocols~\cite{Quach2020, Liu2019, Finkelstein2019, Andolina2018}. However, most existing approaches rely on symmetry arguments and are largely restricted to ensembles of two-level systems.

At the same time, many experimentally relevant quantum platforms—most notably superconducting transmon circuits~\cite{Dou2023, Yang2024, Hu2018}—are intrinsically multilevel systems. Truncating these systems to qubits neglects the internal ladder structure that can, in principle, be exploited to enhance both energy density and lifetime~\cite{Gemme2024}. Despite this, a systematic framework for identifying and exploiting protected energy storage states in interacting multilevel quantum batteries remains lacking~\cite{Rangelov2006}.

In this work, we address this gap by introducing a constructive and thermodynamically consistent method to analyze dissipation in multilevel quantum batteries. Our findings align with recent studies showing that increasing the local Hilbert space dimension (e.g., higher-spin chains) significantly enhances the charging energy and power of interacting quantum batteries~\cite{Sun2025}. Our approach combines the Davies master equation, which ensures correct thermalization behavior, with a Morris--Shore (MS)–type decomposition~\cite{Morris1983} applied to Bohr-frequency–resolved dissipative coupling blocks. This framework allows us to classify excited states according to their dissipative role and to identify not only dark states, but also a hierarchy of excited states that act as~\emph{dissipative funnels}, decaying exclusively into protected manifolds.

In the broader context of controlling energy flow and dissipation in open quantum systems, it is crucial to distinguish between reciprocal and nonreciprocal regimes. Recent studies demonstrate that introducing nonreciprocity can significantly alter the thermodynamics and efficiency of quantum batteries~\cite{PRL_132_210402_2024, PRXEnergy_5_023003_2026}. Our study, in contrast to classical 'reciprocal' dissipative channels, identifies 'funnel' channels that create a geometrically structured dissipative pathway in a reciprocal passive-bath model through MS decomposition. While standard dissipative models predict the thermal equilibrium of energy with the environment, our described 'funnel' mechanism allows for the selective channeling of energy into 'dark' manifolds under dissipative interaction.

We apply this Davies-MS framework to a minimal yet nontrivial model: two interacting qutrit quantum batteries coupled to a common bath. This model captures essential features of superconducting transmon platforms, including ladder-dependent transition strengths and weak anharmonicity. We analytically construct dark, bright, and funnel states, derive quantitative robustness conditions governed by the ratio of interaction strength to anharmonicity, and validate our predictions through numerical simulations. Beyond passive dark-state storage, we propose a battery design strategy in which high-energy funnel states are actively used as storage targets. When combined with selective suppression of specific decay channels, these states can store more energy than the corresponding dark states while maintaining long lifetimes.

The rest of the work is organized as follows: We discuss the quantum battery model and
the Davies master equation in Sec.~\ref{section 2} and present the MS transformation and
singular value decomposition in Sec.~\ref{section 3}. We determine the optimal charging protocol
in Sec.~\ref{section 4}. Finally, we conclude our findings in Sec.~\ref{section 5}.

\section{Model and Davies Master Equation \label{section 2}}

Each quantum battery is modeled as a weakly anharmonic three-level system (qutrit) with eigenenergies~\cite{Blok2021}
\begin{equation}
E_n = n\omega - \frac{\alpha}{2} n(n-1),
\qquad n=0,1,2,
\label{eq:En} 
\end{equation}
where $\omega$ is
the transition frequency in the absence of anharmonicity (i.e., $\alpha$ = 0), and $\alpha$
characterizes the strength of the anharmonicity. As a typical physical system with such anharmonic three-level structure, we consider superconducting transmon qutrits for which
$\alpha>0$ and $\alpha \ll \omega$; hence the upper energy gap $E_2-E_1=\omega-\alpha$ is smaller than the lower one $E_1-E_0=\omega$.

For a transmon qutrit, the lowering operator truncated to the three lowest levels can be written as~\cite{Goss2022, Morvan2021}
\begin{equation}
a = \ket{0}\!\bra{1} + \sqrt{2}\ket{1}\!\bra{2}.
\label{eq:a_def}
\end{equation}

We consider a system of two identical quantum batteries, coupled via an exchange interaction, whose  Hamiltonian reads
\begin{equation}
H_S =
\sum_{j=A,B} \sum_{n=0}^2 E_n \ket{n}_j\!\bra{n}
+ J \left(a_A^\dagger a_B + a_A a_B^\dagger\right),
\label{eq:HS}
\end{equation}
where $J$ is the coupling strength. For transmon qutrits, such an interaction can be realized by capacitive coupling under the rotating wave approximation (RWA)~\cite{Yang2024}.
Additional indirect coupling could be mediated by using a transmission line resonator coupled to both the quantum batteries, which we do not consider here. The RWA reduces the capacitive coupling term $(S_A^- - S_A^+)(S_B^- - S_B^+)$, to the exchange interaction term $J(a_A^\dagger a_B + a_A a_B^\dagger)$. The condition for the RWA is given by $\text{min}\{\omega,\omega-\alpha\}\gg J$.

For charging the batteries, we consider classical drive fields such that the charging Hamiltonian can be written as 
\begin{equation}
    H_{d}(t)=i\Omega_R(t)(a_A^\dagger-a_B^\dagger)+\mathrm{H.c.},
\end{equation}
where $\Omega_R(t)$ is the Rabi frequency associated with the transitions. The choice of an antisymmetric (out-of-phase) drive is not arbitrary but due to an optimization, which will be explained in Sec.~\ref{section 4}.

The batteries are coupled to a common electromagnetic environment through
\begin{equation}
H_{SB}=[(a_A+a_B)+(a_A^\dagger+a_B^\dagger)]\otimes B,
\label{eq:HSB}
\end{equation}
where $B$ stands for bath operators.

For a weakly coupled bath, we employ Davies' construction to write the Gorini-Kossakowski-Sudarshan-Lindblad (GKSL) master equation~\cite{Davies1974,Breuer2002}
\begin{equation}
\dot{\rho}
=-i[H_S+H_d(t),\rho]+\mathcal{D}_{\mathrm D}[\rho],
\label{eq:GKLS}
\end{equation}
where, assuming a low temperature ($T\rightarrow 0$), the dissipator has a contribution only from spontaneous emission described by
\begin{eqnarray}
\mathcal{D}_{\mathrm D}[\rho]
=
\sum_{\Omega>0}\Gamma(\Omega) \qty (
A(\Omega)\rho A^\dagger(\Omega)
-\frac12\{A^\dagger(\Omega)A(\Omega),\rho\}
).
\label{eq:DaviesZeroT}
\end{eqnarray}
Here,
\begin{equation}
A(\Omega)=\sum_{\epsilon'-\epsilon=\Omega}\Pi_\epsilon\, (a_A+a_B)\,\Pi_{\epsilon'},
\label{eq:AOmega}
\end{equation}
with $\Pi_\epsilon$ is the spectral projector of $H_S$ onto energy $\epsilon$.  The term $\Gamma(\Omega)$ in Eq. ~\ref{eq:DaviesZeroT} represents the microscopic interaction of the system with its environment. In our numerical analysis, a constant damping ratio of $\gamma = 0.1$ was assumed for all transitions to reduce complexity.

The Davies jump operators are constructed from the eigenbasis of the undriven interacting system Hamiltonian $H_S$. The charging field $H_d(t)$ is then added as an external coherent control term in the Schrödinger-picture master equation. This separation allows us to identify dark, bright, and funnel states of the interacting open system independently of the specific charging protocol. It is justified provided the drive does not substantially modify the Bohr-frequency structure entering the Davies decomposition. It holds when the drive varies slowly on the reservoir memory timescale. A sufficient
condition is
\begin{equation} \label{eq:markov_con}
|\Omega_R(t)|\,\tau_B \ll 1,
\qquad
|\dot\Omega_R(t)|\,\tau_B^2 \ll 1 ,
\end{equation}
where $\tau_B$ is the bath correlation time. In practice, this requires the
drive modulation timescale
\begin{equation}
\tau_{\mathrm{mod}}(t)\sim
\frac{|\Omega_R(t)|}{|\dot\Omega_R(t)|}
\end{equation}
to be long compared with the dissipative timescale $\Gamma^{-1}$.

\section{Morris-Shore Transformation and Singular Value Decomposition (SVD) \label{section 3}}

The Davies generator decomposes the system--bath coupling into
Bohr-frequency--resolved jump operators $A(\Omega)$, each connecting
eigenstates of $H_S$ whose energies differ by $\Omega$. For a fixed
$\Omega$, the operator $A(\Omega)$ therefore defines a linear map
between two excitation manifolds: for instance, an upper manifold $\mathbf H^{(2)}$
and a lower manifold $\mathbf H^{(1)}$.

Choosing orthonormal bases
$\{\ket{\psi_j^{(A)}}\}$ in the upper manifold and
$\{\ket{\phi_i^{(B)}}\}$ in the lower manifold, the jump operator can
be written as

\begin{equation}
A(\Omega)=
\sum_{i,j}
M_{ij}
\ket{\phi_i^{(B)}}\bra{\psi_j^{(A)}},
\end{equation}
where the matrix elements are obtained directly from the jump operator
as
\begin{equation}
M_{ij}
=
\bra{\phi_i^{(B)}}A(\Omega)\ket{\psi_j^{(A)}} .
\end{equation}

Thus, the matrix $M$ is simply the representation of the Davies jump
operator between two excitation manifolds. It contains the dissipative
transition amplitudes from the upper manifold to the lower one.

The mathematical mechanism underlying the MS
transformation is the singular value decomposition (SVD) in linear algebra.
Traditionally used in closed quantum systems to reduce complex coherent
drives to independent two-level subsystems, this method~\cite{Morris1983} has been extended to open quantum systems in the
framework we present and applied to dissipative coupling blocks
decomposed to Bohr frequencies~\cite{Saadati2016, Kim2015, Finkelstein2019}.

When SVD is applied to the jump matrix $M$ of dimensions
$N_B \times N_A$, it is factorized as

\begin{equation}
M = U \Sigma V^\dagger
\end{equation}

In this separation, the physical meanings of the matrices match as follows: 

The columns of the unitary matrix $V$ define orthonormal linear combinations of states within the upper manifold, representing the effective \textit{source states} that participate in dissipative transitions. Correspondingly, the columns of the matrix $U$ characterize the orthonormal basis of the lower manifold, acting as the designated \textit{target states} for these dissipative channels. Finally, the non-negative diagonal elements $\sigma_i$ of the matrix $\Sigma$ quantify the effective coupling strengths of each independent decay channel, thereby providing a complete characterization of the dissipative landscape. We note that this MS-type decomposition provides a natural basis to separate collective dissipative channels, a methodology sharing conceptual similarities with approaches used to distinguish local and global master equations in quantum thermal machines~\cite{Hofer2017}. It is worth underscoring that this MS-type formulation serves as a foundational notation and stands as a principal result of our paper. The key consequence of this decomposition is that each singular vector pair defines an independent dissipative pathway connecting one rotated
upper state to one rotated lower state. If a singular value is exactly
zero ($\sigma_i=0$), the corresponding upper-state combination is
annihilated by the jump operator,

\begin{equation}
M v_i = 0 ,
\end{equation}
and therefore does not decay through this dissipative block. These null
vectors define the spectator states of the MS transformation.
When such states are also eigenstates of $H_S$, they become exact dark
states of the open-system dynamics~\cite{Zlatanov2020, Finkelstein2019}. Conversely, singular vectors with nonzero singular values define
dissipative channels. Depending on their decay targets, these states may
act as funnel or bright states, as classified below.

\subsection{Dark, Funnel, Bright and Spectator States\label{section 3A}}

The Davies-MS decomposition leads to a natural classification of
eigenstates based on the structure of the jump operator.

\paragraph{Dark states.}

States annihilated by all relevant jump operators and invariant under
the coherent Hamiltonian. These states do not decay under collective
dissipation and form decoherence-free storage manifolds.

\paragraph{Funnel states.}

Excited states corresponding to nonzero singular values whose decay
targets lie entirely inside the dark subspace of the lower manifold.
These states act as dissipative loading channels, transferring
population irreversibly into protected storage states.

\paragraph{Bright states.}

States whose decay channels connect to radiative sectors that eventually
lead to the ground state. These states rapidly lose stored energy and
are unsuitable for battery operation.

\paragraph{Spectator states.}

Null vectors of a dissipative block that are not eigenstates of the
system Hamiltonian. When hybridization with $H_S$ is weak, these states
behave as approximate dark states.

In contrast to qubit batteries, multilevel ladder structure enables
funnel states at energies higher than the dark states themselves.


\subsection{Explicit Determination of Dark, Funnel, and Bright States \label{section_3_b}}

\begin{figure}[H]
\centering
\begin{tikzpicture}[
    x=1cm,y=1cm,
    level/.style={thick},
    decay/.style={-{Latex[length=2mm]}, thick},
    statelabel/.style={font=\fontsize{10}{10}\selectfont}
]

\draw[level] (0.8,4.2) -- (2.4,4.2);
\node[statelabel,anchor=west] at (2.5,4.2) {$\ket{E_+}$};

\draw[level] (1.0,3.6) -- (2.2,3.6);
\node[statelabel,anchor=west] at (2.3,3.6) {$\ket{E_-}$};

\draw[level] (0.8,2.2) -- (2.4,2.2);
\node[statelabel,anchor=west] at (2.5,2.2) {$\ket{B_1}$};

\draw[level] (0.8,0.8) -- (2.4,0.8);
\node[statelabel,anchor=west] at (2.5,0.8) {$\ket{00}$};

\draw[level] (4.5,4.0) -- (6.1,4.0);
\node[statelabel,anchor=west] at (6.2,4.0) {$\ket{A_2}$};

\draw[level] (4.5,2.2) -- (6.1,2.2);
\node[statelabel,anchor=west] at (6.2,2.2) {$\ket{D_1}$};

\draw[decay] (1.55,4.12) -- (1.55,2.3);
\draw[decay] (1.9,3.52) -- (1.9,2.3);
\draw[decay] (1.72,2.12) -- (1.72,0.9);
\draw[decay] (5.3,3.92) -- (5.3,2.3);

\end{tikzpicture}
\caption{Dressed-state decay structure of the two-qutrit system.}
\label{fig:dressed_decay_structure}
\end{figure}

The RWA Hamiltonian~\eqref{eq:HS} conserves the total excitation number
\begin{equation}
N=n_A+n_B,
\end{equation}
so that the Hilbert space decomposes into invariant manifolds $N=0,1,2,3,4$.
The physically most important ones for the present discussion are $N=1$ and $N=2$.

\subsubsection{$N=1$ manifold}

In the basis $\{\ket{10},\ket{01}\}$, the Hamiltonian block is
\begin{equation}
\mathbf H^{(1)}=
\begin{pmatrix}
\omega & J \\
J  & \omega
\end{pmatrix}.
\end{equation}
Hence the exact eigenstates are
\begin{align}
\ket{B_1}&=\frac{1}{\sqrt2}\qty(\ket{10}+\ket{01}),
&
E_{B_1}&=\omega+J,
\\
\ket{D_1}&=\frac{1}{\sqrt2}\qty(\ket{10}-\ket{01}),
&
E_{D_1}&=\omega-J.
\end{align}
Acting with the collective lowering operator
\begin{equation}
L\equiv a_A+a_B
\end{equation}
gives
\begin{equation}
L\ket{B_1}=\sqrt2\ket{00},
\qquad
L\ket{D_1}=0.
\end{equation}
Therefore $\ket{D_1}$ is an \emph{exact dark state} of the collective dissipation and also an eigenstate of $H_S$.

\subsubsection{$N=2$ manifold}
In the bare basis $\{\ket{20},\ket{11},\ket{02}\}$,
\begin{equation}\label{eq:H2bare}
\mathbf H^{(2)}=
\begin{pmatrix}
2\omega-\alpha & \sqrt2 J & 0\\
\sqrt2 J & 2\omega & \sqrt2 J\\
0 & \sqrt2 J & 2\omega-\alpha
\end{pmatrix},
\end{equation}
It is convenient to introduce the symmetric and antisymmetric spectral edge states of the three-level system
\begin{equation}
\ket{S_2}=\frac{1}{\sqrt2}\qty(\ket{20}+\ket{02}),
\quad
\ket{A_2}=\frac{1}{\sqrt2}\qty(\ket{20}-\ket{02}).
\end{equation}
In the basis $\{\ket{A_2},\ket{S_2},\ket{11}\}$, the block becomes
\begin{equation}
\mathbf H^{(2)}=
\begin{pmatrix}
2\omega-\alpha & 0 & 0\\
0 & 2\omega-\alpha & 2J \\
0 & 2J & 2\omega
\end{pmatrix}.
\label{eq:H2SAbasis}
\end{equation}
Thus $\ket{A_2}$ is an \emph{exact eigenstate} of $H_S$ with energy
\begin{equation}
E_{A_2}=2\omega-\alpha,
\end{equation}
while $\ket{S_2}$ hybridizes with $\ket{11}$ into two bright-sector eigenstates
\begin{equation}
\ket{E_\pm}=\cos\theta\,\ket{S_2}\pm \sin\theta\,\ket{11},
\qquad
\tan 2\theta=\frac{4J}{\alpha},
\label{eq:pmstates}
\end{equation}
with eigenenergies
\begin{equation}
E_\pm=2\omega-\frac{\alpha}{2}\pm\frac12\sqrt{\alpha^2+16J^2}.
\label{eq:Epm}
\end{equation}

\subsection{Dissipative MS decomposition}

The decisive object for decay classification is not only $H_S$ but the action of the collective lowering operator $L=a_A+a_B$ between excitation manifolds.

\subsubsection{$N=2\to N=1$ block}

Using the $\{\ket{A_2},\ket{S_2},\ket{11}\}$ basis in $N=2$ and the $\{\ket{D_1},\ket{B_1}\}$ basis in $N=1$, one finds
\begin{align}
L\ket{A_2} &= \sqrt2\ket{D_1},\\
L\ket{S_2} &= \sqrt2\ket{B_1},\\
L\ket{11}  &= \sqrt2\ket{B_1}.
\end{align}
Therefore the dissipative coupling matrix is
\begin{equation}
M_{2\to1}=
\begin{pmatrix}
\sqrt2 & 0 & 0\\
0 & \sqrt2 & \sqrt2
\end{pmatrix},
\label{eq:M21}
\end{equation}
where the rows correspond to $\{\ket{D_1},\ket{B_1}\}$ and the columns to $\{\ket{A_2},\ket{S_2},\ket{11}\}$.

This block immediately yields the correct taxonomy.

\paragraph{Funnel state.}
Because
\begin{equation}
L\ket{A_2}=\sqrt2\ket{D_1},\label{eq:funnel_condition}
\end{equation}
the state $\ket{A_2}$ decays \emph{exclusively} into the protected $N=1$ dark state. It is therefore an exact funnel state.

\paragraph{Spectator state of the dissipative block.}
The null vector of the second row of Eq.~\eqref{eq:M21} is
\begin{equation}
\ket{S_{\mathrm{spec}}^{(2)}}\propto
\sqrt2\ket{S_2}-\sqrt2\ket{11}
\propto
\ket{20}-\sqrt2\ket{11}+\ket{02}.
\label{eq:spectator}
\end{equation}
By construction,
\begin{equation}
L\ket{S_{\mathrm{spec}}^{(2)}}=0.
\end{equation}
Hence this is a spectator state of the $N=2\to1$ dissipative block.

\subsection{Effect of anharmonicity and interaction}

While $\ket{X_{\rm spec}}$ is annihilated by the dissipative block
$M_{2\to 1}$, it is not, in general, an eigenstate of $H_S$ when
$\alpha \neq 0$.
Restricting the Hamiltonian~\eqref{eq:HS} to the $N=2$ manifold yields the Hamiltonian~\eqref{eq:H2bare}, seeking an eigenstate of the symmetric form
\begin{equation}
\ket{\psi} \propto (1,x,1),
\end{equation}
one finds that $x$ satisfies the quadratic equation
\begin{equation}
\sqrt{2}J x^2 - \alpha x - 2\sqrt{2}J = 0.
\label{eq:x_eq}
\end{equation}
For weak anharmonicity $\alpha \ll J$, the physically relevant solution is
\begin{equation}
x \simeq -\sqrt{2} + \frac{\sqrt{2}}{4}\frac{\alpha}{J} + \mathcal{O}\!\left(\frac{\alpha^2}{J^2}\right).
\label{eq:x_approx}
\end{equation}

Eq.~\eqref{eq:x_approx} quantifies the deviation of the exact eigenstate
from the ideal spectator state and provides a robustness criterion:
The overlap with the dark manifold improves as $J/\alpha$ increases.
Therefore, the spectator state
is an exact dark state only in the harmonic limit $\alpha=0$. For weak anharmonicity, it is not exact, but remains an \emph{approximate dark state} when $
\alpha/J \ll 1$
\label{eq:approxdark}
~\cite{Zlatanov2020}.

\paragraph{Bright states.}
The orthogonal combinations in the $\{\ket{S_2},\ket{11}\}$ sector couple to $\ket{B_1}$ and then to $\ket{00}$. These are the bright states. In the exact eigenbasis, they are the two dressed states $\ket{E_\pm}$ of Eq.~\eqref{eq:pmstates}.

\section{Determination of Optimal Charging \label{section 4}}

MS analysis allows us to determine an optimal charging model.
Bright states $\ket{E_{\pm}}$ of the $N=2$ excitation manifold are attractive to charge due to their high energies, but the zero temperature environment couples them to the bright state $\ket B_1$, and hence they are short-lived in open system conditions. The spectator state is also of high energy, but it is not a true dark state of the anharmonic qutrit system. The exact dark state $\ket{D_1}$ of the system is the longest-lived but at a lower energy. Hence, our optimal choice would be the funnel state $\ket{A_2}$.

Provided the drive is turned off after preparation or is engineered not to destabilize the protected sector, the funnel state $\ket{A_2}$ is therefore an optimal charging target, as it sits at higher energy than the dark state $\ket{D_1}$ of the system and its decay is structured so that it feeds the protected dark state $\ket{D_1}$ rather than the bright lossy channel.

For the simple two-qutrit
model, we can analytically identify the charging mechanism leading to the funnel
states. As the bath operator is symmetric, $L=a_A+a_B$, an antisymmetric drive excites the initially ground state $\ket{00}$ to the antisymmetric states in the single-excitation $N=1$ manifold first, then the other antisymmetric states in the higher excitation manifolds.

Here, we consider continuous antisymmetric driving as a proof-of-principle and numerically convenient protocol. Our approach is prone to leakage into higher excitation sectors and to imperfect charging with low fidelity to the funnel and dark state. Alternatively, one could use a pulse protocol to achieve higher-fidelity charging into the funnel state.
Another alternative, which is also scalable for a larger number of qutrits or multi-level systems (qudits), is constructed in the next section.

\subsection{Funnel-state targeting and scalable charging protocol}

The continuous antisymmetric driving considered above provides a simple
proof-of-principle charging mechanism, but is not optimal. Because the drive
acts simultaneously on multiple transitions, population leaks into higher
excitation manifolds and bright sectors, leading to imperfect loading of the
desired funnel state. A more consistent strategy is to first identify the
funnel states of the interacting open system and then design a charging
protocol that selectively transfers population into those states.

This viewpoint generalizes naturally to larger arrays of qutrits or, more
generally, multilevel qudits.

\subsubsection{Identification of funnel states}

Let $\{\ket{\phi_k}\}$ denote eigenstates of the interacting Hamiltonian $H_S$,
\begin{equation}
H_S \ket{\phi_k}=E_k\ket{\phi_k}.
\end{equation}

The dissipative structure is determined by the Davies jump operators, which were given in Eq.~(\ref{eq:AOmega}) from which one constructs the decay graph between eigenstates.

A state $\ket{\phi_k}$ is classified as

\begin{itemize}
\item dark if $A(\Omega)\ket{\phi_k}=0$ for all $\Omega$,
\item funnel if the action of $A(\Omega)$ maps it entirely into
the protected dark manifold,
\item bright if it decays into radiative sink states.
\end{itemize}

This classification is independent of the charging protocol and depends only
on the interacting Hamiltonian and the system–bath coupling.

\subsubsection{Targeted funnel-state charging}

Once the funnel states are identified, the charging problem reduces to
preparing a chosen funnel state. Let $\ket{\phi_F}$ denote a selected funnel
state. The objective is to design a control Hamiltonian
\begin{equation}
H_{\mathrm{ch}}(t)=\sum_\ell u_\ell(t) H_\ell
\label{H_charging}
\end{equation}
such that
\begin{eqnarray}
U(T)\ket{\psi_0}&\approx& \ket{\phi_F},
\nonumber\\
U(T)&=&\mathcal{T}\exp\left[
-i\int_0^T (H_S+H_{\mathrm{ch}}(t))dt
\right],
\end{eqnarray}
where $\ket{\psi_0}$ is the initial ground state. In Eq.~(\ref{H_charging}), the $u_l(t)$ parameters represent the time-dependent laser pulses that we apply externally. $H_l$ are the physical transitions that we can manipulate in the system. 

The optimization target is to maximize the fidelity with the funnel state,
\begin{equation}
\max_{u_\ell(t)} 
\; \mathcal{F}
=
\bra{\phi_F}\rho(T)\ket{\phi_F},
\label{eq:fidelity_target}
\end{equation}
i.e., the overlap between the final state $\rho(T)$ and the funnel state $\phi_F$~\cite{Brif2010}. The optimization of the control fields $u_l(t)$ to maximize the fidelity $\mathcal{F}$ can also be performed using gradient-based algorithms such as GRAPE~\cite{Khaneja2005}, which are well-suited for high-dimensional multilevel systems.

After the charging pulse is turned off, dissipation drives the system toward
the protected dark manifold,
\begin{equation}
\ket{\phi_F}
\longrightarrow
\text{dark manifold}.
\end{equation}

Thus, the funnel state acts as a high-energy loading state, while the dark
subspace provides long-term storage.

\subsubsection{Selection among multiple funnel states}

If multiple funnel states exist, one may compare them using their stored energy
\begin{equation}
E_F=\bra{\phi_F}H_S\ket{\phi_F},
\end{equation}
and we define an effective decay rate $\Gamma_F$ for each funnel state to quantify its stability
\begin{equation}
\Gamma_F=
\sum_{\Omega>0}\gamma(\Omega)
\bra{\phi_F}A^\dagger(\Omega)A(\Omega)\ket{\phi_F}.
\end{equation}

The optimal target is typically the highest-energy funnel state whose decay
graph terminates in the protected dark manifold. Importantly, this comparison
is performed \emph{after} identifying funnel states, rather than searching
over arbitrary eigenstates.

\subsubsection{Scalable recipe}

The proposed scalable protocol for efficient charging of quantum batteries follows these steps: First, the system Hamiltonian ($H_S$) is diagonalized to generate Davies jump operators decomposed into Bohr frequencies. Next, the SVD-based MS transform is applied to the dissipative matrix, which represents the transitions between excitation manifolds, to determine the new bases and decay channels of the system. Through this separation, \textit{dark states} (zero singular values), unaffected by the bath, and \textit{funnel states} that direct energy directly to protected subspaces are identified. In the final stage, a laser pulse is applied to prepare the selected funnel state with high fidelity, ensuring that the system subsequently relaxes toward the protected dark manifold when the drive is switched off, allowing for long-lasting dark states. This procedure separates the identification of protected energetic states from the control design and scales naturally to larger qudit arrays.

Dark states provide maximal protection against dissipation but typically occur
in lower excitation manifolds and therefore store limited energy. Funnel states,
in contrast, lie in higher-energy sectors while decaying exclusively into the
protected dark manifold. Consequently, they are metastable rather than strictly
dark, but their decay loads long-lived storage states. From a quantum battery
perspective, this is advantageous: charging into a funnel state stores more
energy initially, while dissipation subsequently transfers population into the
protected dark manifold. The relevant figure of merit, therefore, balances stored
energy and lifetime, and funnel states generally outperform direct dark-state
charging due to their higher excitation energy, combined with structured
dissipative protection.

\subsubsection{Two-qutrit example}

For the two-qutrit model discussed above, the antisymmetric two-excitation
state $\ket{F_2}$ acts as the funnel state, while the antisymmetric
single-excitation state $\ket{D_1}$ forms the protected dark manifold.
The charging protocol therefore aims to prepare $\ket{F_2}$, after which
dissipation loads the long-lived storage state $\ket{D_1}$.

In our approach, the dark state condition is defined with respect to the collective bath
coupling operator, not with respect to the charging Hamiltonian. Consequently,
the one-excitation dark state \(\ket{D_1}\) can still be coherently accessed by
a suitably chosen antisymmetric drive. The reason for not targeting
\(\ket{D_1}\) directly is therefore not poor controllability, but limited
storage capacity: \(\ket{D_1}\) lies in a lower excitation manifold than the
funnel state \(\ket{F_2}\) and stores less energy. The funnel state is thus the
preferred charging target because it combines higher stored energy with a
structured dissipative pathway into the protected dark manifold. In this sense,
the charging strategy is not ``dark-state preparation'' but rather
``high-energy funnel-state preparation followed by dissipative loading into the
dark storage sector''.

We restrict the charging field to a minimal single-envelope
ansatz, in which the same Rabi amplitude drives both local \(0\leftrightarrow1\)
and \(1\leftrightarrow2\) transitions with fixed relative matrix elements inherited
from the transmon ladder operator. Within this restricted control family, the
physically relevant optimization variables are the relative phase between the two
local drives, the pulse amplitude, and the pulse duration. The optimization does
not generate a qualitatively different charging mechanism; rather, it identifies
the antisymmetric configuration as the optimal one for loading the funnel sector,
while the pulse area determines the best transient fidelity to the target state.
A more selective charging protocol would require transition-resolved controls for
the \(0-1\) and \(1-2\) transitions separately, enabling sequential
\(\ket{00}\to\ket{D_1}\to\ket{F_2}\) population transfer with reduced leakage
, which we leave for future work.

\subsection{Possible experimental realization}
The minimal model considered here is naturally motivated by superconducting
transmon circuits operated beyond the computational qubit subspace~\cite{Koch2007, Gu2017}. The three lowest transmon levels provide a weakly anharmonic ladder
$\{\ket{0},\ket{1},\ket{2}\}$ with transition frequencies $\omega_{01}$ and $\omega_{12}=\omega_{01}-\alpha$, which is precisely the
local qutrit structure assumed in Eq.~\ref{eq:En}. Experimental superconducting-qutrit processors have already demonstrated addressable qutrit levels, single-qutrit control, qutrit randomized benchmarking, and two-qutrit operations~\cite{Blok2021,Morvan2021,Goss2022}.
The coherent exchange term $J(a_A^\dagger a_B+a_A a_B^\dagger)$ can arise from capacitive coupling between neighboring transmons in the rotating-wave regime, provided $J\ll\omega_{01},\omega_{12}$. The antisymmetric charging Hamiltonian used in our simulations corresponds experimentally to microwave drives applied to the two qutrits with a relative phase of $\pi$. More selective implementations could use transition-resolved pulses on the $0\leftrightarrow1$ and $1\leftrightarrow2$ transitions, as commonly employed in qutrit-control
experiments.
The collective decay channel assumed in the Davies construction requires the two qutrits to couple predominantly to a common electromagnetic environment, for example a shared resonator, transmission line, or waveguide~\cite{Lalumiere2013}. In that case, the radiative system operator is proportional to $a_A+a_B$, giving the collective lowering operator used in Eq.~\ref{eq:HSB}. Local relaxation and dephasing
would reduce the lifetime of the protected component, but the dark--bright--funnel classification remains the relevant starting point whenever the collective radiative channel is the dominant decay pathway~\cite{Chang2018}.

The required ingredients are therefore within the scope of present
superconducting-qutrit technology. Experiments have demonstrated weakly
anharmonic three-level transmon ladders, coherent qutrit control,
two-qutrit operations, and qutrit-level benchmarking. Our model assumes
the standard transmon hierarchy
\(\omega_{01},\omega_{12}\gg J,\Omega_R,\gamma\), together with a
dominant collective radiative channel. These assumptions are consistent
with superconducting-circuit architectures in which qutrit transitions
are individually addressable and multiple artificial atoms are coupled
to a shared electromagnetic environment.
\vspace{2em}

\section{Results and Discussion \label{section 5}}


Numerical simulations of the charging and discharging phases are performed using the Davies-type Markovian master equation derived in Section~\ref{section 2} (Eq.~\ref{eq:GKLS}).

\subsection{Energy storage dynamics}

We first examine the dissipative energy storage dynamics by calculating 
\begin{equation}
\Delta E(t)=\mathrm{Tr}[H_S\rho(t)]-E_{\mathrm{gs}}\label{average_en}
\end{equation}
starting from the ground state, following an optimized charging protocol. By keeping the initial state fixed, we analyze how the system's anharmonicity ($\alpha$) influences its ability to naturally populate protected states without the need for explicit dark-state preparation.

\begin{figure}[H]
	\centering
	  \includegraphics[width=0.95\linewidth]{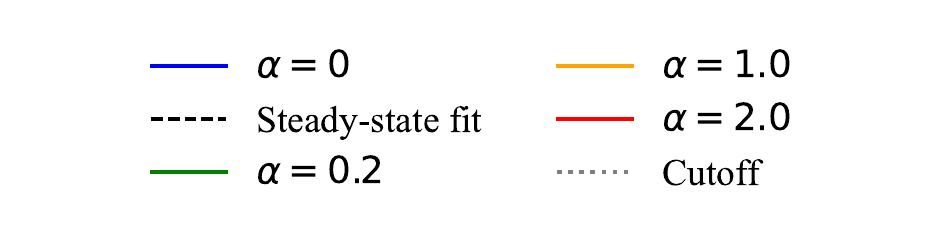}
	\subfloat[]{
		\label{fig:two_qutrit_a}
		 \includegraphics[width=0.95\linewidth]{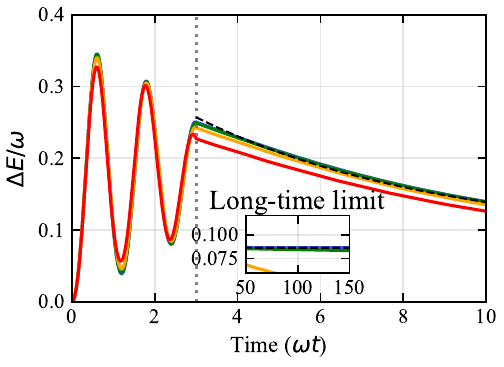}}

	\subfloat[]{
		\label{fig:two_qutrit_b}
		\includegraphics[width=0.95\linewidth]{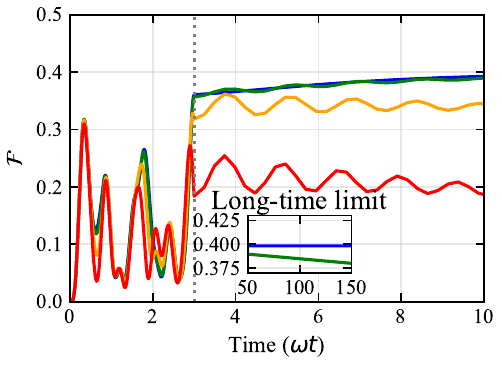}}
	
	\caption{
		Energy storage dynamics (a) and fidelity to the $N=2$ MS spectator reference state (b) for the two-qutrit system under various anharmonicity ($\alpha$) levels. Energy and time are measured in dimensionless units. The charging field is turned off at the exact moment ($t_c\approx3.0/\omega$) the population of the target spectator state reaches its global maximum during the evolution of the harmonic system ($\alpha=0$). The dashed black curve in (a) represents an asymptotic exponential fit to the post-cutoff data for the ideal harmonic limit ($\alpha=0$), strictly confirming that the exact dark state dynamics stabilize to a finite steady-state energy storage. In panel (b), fidelity to the $N=2$ spectator state ($\approx 0.40$ for the blue and green) is dynamically generated and persists at long times, explaining the residual stored energy in panel (a). The insets show the long-time behavior of the steady-state dynamics. Parameter values are set to $J=1.0$, $\gamma=0.1$, and $\omega=10.0$.
	}
	\label{fig:two_qutrit}
\end{figure}

The spectator-state fidelity approaches a finite plateau after the charging stage in 
Fig.~\ref{fig:two_qutrit}. The stored energy Fig.~\ref{fig:two_qutrit_a} and the fidelity with respect to the $N=2$ MS spectator reference state Fig.~\ref{fig:two_qutrit_b} are shown for two-qutrit batteries driven by an antisymmetric continuous-wave field. Upon activation of the charging field, the system exhibits coherent oscillations associated
with population transfer across excitation manifolds. As shown in Fig.~\ref{fig:two_qutrit_a}, the stored
energy $\Delta E(t)$ increases rapidly during the charging phase and reaches a maximum value
determined by the accessible excitation sectors. The subsequent dynamics after the
drive is switched off are governed solely by dissipative processes described by the Davies
master equation (Eq.~\ref{eq:GKLS}).
At this stage, the antisymmetric drive not only maintains the system at $N=1$ but also, because of the multilevel ladder structure, facilitates access to higher excitation manifolds, leading to a significant increase in the stored energy by activating funnel states in manifolds at $N=2$ (and above).

The absence of an initial lag in the energy uptake is attributed to the direct coherent drive causing immediate excitation of accessible sectors, while
subsequent dissipative pathways transfer the appropriate component into the protected manifold. Furthermore, the initial transient oscillations observed in both the stored energy and the spectator state fidelity originate from the interplay between the coherent charging field and the dissipative environment. During the early stages of charging, the coherent drive induces underdamped Rabi-like oscillations between the ground state and the funnel states. These coherent population fluctuations are directly reflected in the total energy and modulate the instantaneous decay rate into the dark manifold, causing the ripples in $\mathcal{F}(t)$ before dissipation fully stabilizes the system into the steady-state plateau.

   In Fig.~\ref{fig:two_qutrit_a}, the fact that the stored-energy curves for different values of $\alpha$ relax to distinct post-charging plateaus highlights the critical impact of anharmonicity on energy retention. Ideal (harmonic) or low-harmonic transmon qutrits maximize the asymptotic energy because they more easily preserve access to higher levels. When anharmonicity becomes too high, symmetry is broken, disrupting the structure of dark-spectator/funnel states, and energy is dissipated more efficiently into the bath, settling on lower-level plateaus. This behavior is consistent with a suppression of higher-level population due to increased anharmonicity, which effectively limits transitions into higher excitation manifolds~\cite{Massa2025}.
   \\
   
  Fig.~\ref{fig:two_qutrit_b} shows that the spectator-state fidelity approaches a finite plateau after the charging stage. The reason it cannot reach exactly perfect (unit) fidelity is that the continuous drive is not transition-selective, it populates several excitation sectors, preventing unit fidelity with the target spectator state; therefore, some of the energy is necessarily distributed to other excitation sectors. However, the fact that fidelity remains stable on a horizontal line confirms that the battery has established a stable, non-decaying reservoir of dark states for $\alpha=0$ and only an approximate dark state when $\alpha\neq0$.

 The post-cutoff oscillations of the target-state fidelity in
Fig.~\ref{fig:two_qutrit_b} have a distinct origin from the dissipative relaxation of the stored energy. The reference state in the two-excitation sector contains components in the $|20\rangle$, $|11\rangle$, and $|02\rangle$ configurations. For $\alpha \neq0$, the anharmonic contribution shifts these components unequally, so that the ideal
MS spectator-state combination is not, in general, an exact eigenstate of $H_S$. After the charging field is switched off, its dressed-state components therefore acquire different dynamical phases, producing coherent beating in the fidelity. In contrast, the corresponding energy
expectation value is conserved under the unitary evolution generated by $H_S$ after the cutoff; its remaining time dependence is consequently governed mainly by dissipative relaxation and is much smoother.

\subsection{Multilevel advantage: qutrit versus qubit batteries}

A central motivation for considering multilevel quantum batteries is the
possibility of increasing the stored energy per constituent without
sacrificing robustness against dissipation.
To assess this advantage, we compare the two-qutrit battery with a qubit
benchmark under an otherwise identical system--bath coupling and interaction
strength.

\begin{figure}[H]
	\centering
	 \includegraphics[width=0.95\linewidth]{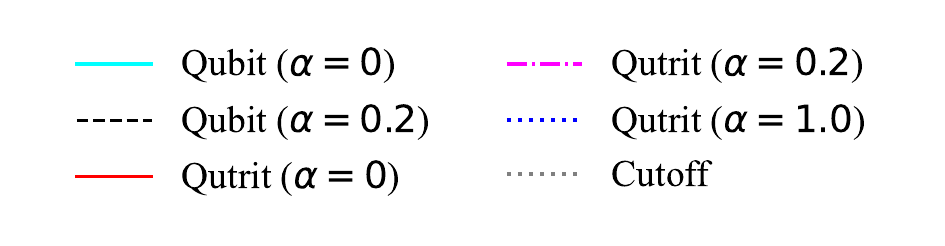}
	\subfloat[]{
		\label{fig:qubit_qutrit_a}
		 \includegraphics[width=0.95\linewidth]{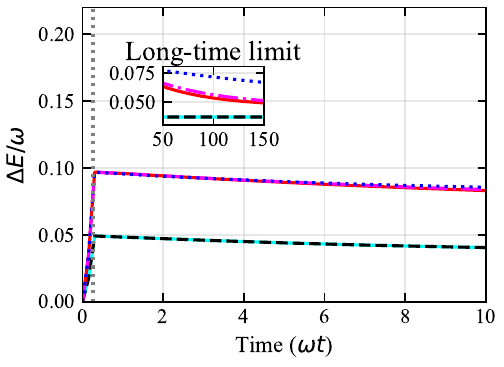}
	}

	\subfloat[]{
		\label{fig:qubit_qutrit_b}
		\includegraphics[width=0.95\linewidth]{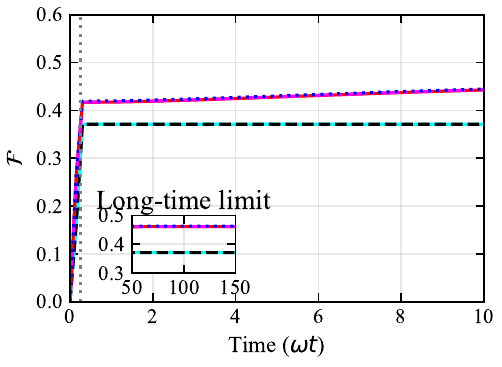}
	}
	
	\caption{
		Performance comparison between qubit and qutrit battery systems under an antisymmetric drive. (a) Stored energy versus time and (b) fidelity to the $N=1$ exact dark state versus time are given in dimensionless units. The vertical gray dotted line indicates the universal charger cutoff time ($t_c \approx 0.26 / \omega$) for all curves, which is calibrated specifically for the harmonic qutrit scenario ($\alpha=0$) to maximize its dark state population. The insets show the long-time behavior of the steady-state dynamics. Other parameters are $J=1.0$, $\gamma=0.1$, and $\omega=10.0$.
	}
	\label{fig:qubit_qutrit}
\end{figure}

We show the stored energy in Fig.~\ref{fig:qubit_qutrit_a} and dark-state fidelity in Fig.~\ref{fig:qubit_qutrit_b}
for two-qubit and two-qutrit batteries, following the same charging-and-relaxation
protocol. However, the charging process in this instance has excited the system to the $N=1$ manifold, ensuring a fair comparison between the qubit and the qutrit. For the qubit benchmark, the anharmonicity parameter is irrelevant; consequently, the two qubit curves overlap~\cite{Elghaayda2025}. However, for the qutrit case, we see the difference between them. The results for $\alpha = 1$, $\alpha = 0.2$, and the harmonic limit $\alpha=0$ are ordered according to their anharmonicity levels. This behavior arises because increased anharmonicity acts as an effective barrier that suppresses population transfer to the upper states of the $N=2$ manifold, thereby preventing the system from fully exploiting the multilevel structure. Consequently, this effective level blockade reduces the overall storage capacity toward the two-level qubit limit~\cite{Massa2025}. 

The steady-state stored energy in the qutrit system exceeds that of the qubit benchmark by a factor of approximately 1.7-1.8 under identical coupling conditions, as can be seen from the inset given in Fig.~\ref{fig:qubit_qutrit_a}. This enhancement arises from access to higher excitation manifolds enabled by the multilevel
structure. This increase originates from an increased effective Hilbert space that allows population
to be distributed across higher-energy states before dissipative relaxation. The oscillations
in the fidelity in Fig.~\ref{fig:two_qutrit_b} are absent in the $N = 1$ sector relevant to Fig.~\ref{fig:qubit_qutrit}. Since the local
anharmonicity term is proportional to $n(n-1)$, it vanishes for occupations $n = 0, 1.$ The antisymmetric state $|D1\rangle = (|10\rangle-|01\rangle)/ \sqrt2$ therefore remains an exact eigenstate of $H_S$ for arbitrary $\alpha$, accumulating only a global phase after the charger cutoff. Its fidelity consequently does not show post-cutoff coherent oscillations.

The asymptotic fidelity saturates at approximately 0.45, consistent with partial population transfer into the protected subspace under continuous driving and dissipative
dynamics. This highlights that the multilevel advantage in the present setting is not enhanced charging speed or stronger coherence, but increased energy density under dissipative evolution. Importantly, this enhancement arises without invoking additional symmetries or
fine-tuned control, and is naturally captured by the Davies-MS
analysis.

It is crucial to highlight the ultimate effect of anharmonicity on the storage dynamics of a quantum battery. Anharmonicity alters battery output based on total charging time. As $\alpha \to \infty$, an energy level blockade occurs. The $\ket{2}$ state receives an energy penalty, and the drive stops interacting with these upper states; consequently, the storage capacity of the qutrit battery drops toward that of the two-level qubit benchmark, as evidenced by the degraded plateaus in Fig.~\ref{fig:two_qutrit}. However, in the ultra-short charging regime, the \(\alpha=1.0\) curve in Fig.~\ref{fig:qubit_qutrit} shows a larger trapped energy at the selected cutoff time. This should be interpreted as a transient timing effect: anharmonic detuning changes the phase of the Rabi-like oscillations and can suppress population return to the ground state at that particular cutoff time.

Another notable feature in the post-cutoff fidelity dynamics in both Fig.~\ref{fig:two_qutrit_b} and Fig.~\ref{fig:qubit_qutrit_b} is the slight but continuous post-cutoff rise observed exclusively for qutrit batteries. During the rapid charging phase, the multi-level ladder structure of qutrits allows for the transient population of higher excitation manifolds ($N \ge N_{target}$). Once the drive is abruptly switched off, the subsequent cascade decay from these higher bright states is effectively funneled into the respective target spectator-dark states. This delayed dissipative feeding mechanism, which is fundamentally limited in strict two-level qubit systems, further enhances the steady-state fidelity and demonstrates a distinct structural advantage of utilizing $d > 2$ artificial atoms for dark-state energy storage.

Our numerical solver explicitly splits the dynamics into a driven phase and a purely dissipative phase post-cutoff. The perfectly horizontal plateaus observed for $t > t_c$ confirm that once the external drive is removed, the populated dark state lies entirely within the kernel of the dissipator ($\mathcal{L}[\rho_{\text{dark}}] = 0$), completely halting any further energy exchange or population leakage.

The relaxation dynamics after the charging stage exhibit a clear separation of timescales:
an initial transient decay followed by convergence to a finite asymptotic energy. This
long-time behavior reflects the presence of protected subspaces that inhibit complete
relaxation to the ground state. The apparent difference between the fidelity and energy curves after the charger cutoff originates 
from the fact that they quantify different parts of the state. The fidelity in Fig.~\ref{fig:two_qutrit_b} measures the 
projection of the evolved density matrix onto the chosen protected target state or manifold. 
Once this projected component has been generated, it remains approximately stationary because 
it lies in, or close to, the kernel of the collective dissipator. By contrast, the stored energy in 
Fig.~\ref{fig:two_qutrit_a} is computed from the full density matrix as defined in Eq.~\ref{average_en} and therefore includes population in bright and other nonprotected components populated during the 
charging stage. After the drive is switched off, these nonprotected components continue to relax through 
the Davies dissipator, reducing the total stored energy, while the projection onto the protected target 
subspace remains nearly constant. Thus the saturated fidelity and the decreasing energy are not 
contradictory: they reflect the coexistence of a protected state component and residual unprotected 
population in the full post-charging state.

\subsection{Qubit benchmark and connection to subradiance}

For the qubit benchmark, the observed behavior can be interpreted in terms of
well-known superradiant and subradiant decay channels~\cite{Dicke1954}.
Antisymmetric states are protected against collective emission, leading to
long-lived energy storage.

In the multilevel case, however, protected behavior is no longer restricted to
the single-excitation sector.
The internal ladder structure of qutrits enables spectator and dark states in
higher excitation manifolds, as well as states that decay preferentially into
protected subspaces.
This qualitative difference underlies the enhanced storage capacity of
multilevel batteries and has no direct analogue in qubit systems.

\subsection{Robustness against anharmonicity}

Weak anharmonicity, relevant for transmon platforms, breaks exact dark-state conditions. Analytically, we find that the deviation of the optimal dark manifold from the exchange-symmetric subspace scales as $\mathcal{O}(\alpha/J)$, where $\alpha$ is the anharmonicity. Numerical simulations confirm that increasing the interaction strength restores approximate dark behavior.

\begin{figure}[H]
    \centering
     \includegraphics[width=0.95\linewidth]{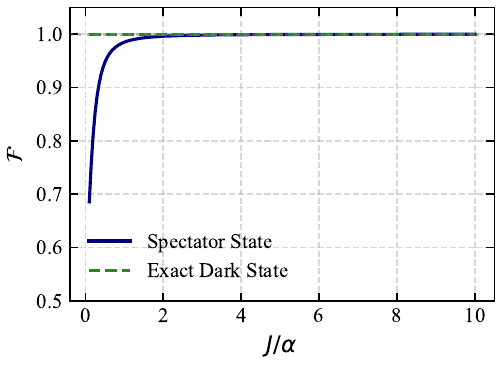}
    \caption{ Robustness of the approximate dark states as a function of interaction strength under strong anharmonicity ($\alpha = 1.0$). The plot shows the overlap between the MS spectator state and the true dark eigenstate of the anharmonic Hamiltonian as a function of the dimensionless ratio $J/\alpha$. Increasing interaction strength relative to anharmonicity restores the approximate dark behavior, consistent with the analytic scaling derived in Eq.~(\ref{eq:x_approx}). All simulations are performed with a system-environment coupling rate $\gamma = 0.1$.}
    \label{fig:robustness}
\end{figure}

\subsection{Funnel states as dissipative intermediates}

\begin{figure}[H]
    \centering
      \includegraphics[width=0.95\linewidth]{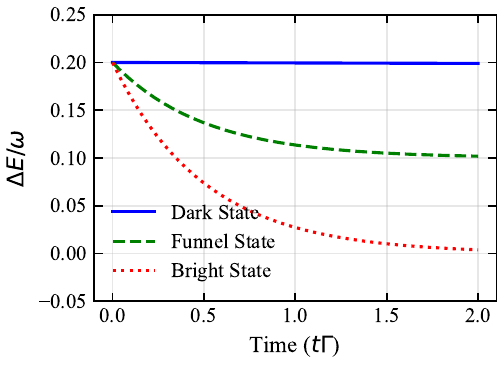}
    \caption{Decay dynamics of different initial states under an anharmonicity $\alpha=0.2$. The horizontal axis is scaled by the effective decay rate $\Gamma$, where we assume a uniform system-environment coupling $\gamma = 0.1$ for all numerical simulations, and energy is dimensionless. The dark state exhibits a longer lifetime than other states. The funnel state decays to the $N=1$ manifold dark state and becomes steady there. The bright state decays rapidly to the ground state, losing all its energy. Other parameters are $J=1.0$, and $\omega=10.0$.}
    \label{fig:initialstates}
\end{figure}

Funnel states play a distinct dynamical role as intermediate-energy states that decay exclusively into protected dark subspaces. As illustrated in Fig.~\ref{fig:initialstates}, these states initially store a relatively large amount of energy and subsequently transfer population into long-lived dark states under dissipative evolution.

This behavior establishes a separation between energy storage and long-term protection: funnel states provide access to higher-energy sectors, while the dark manifold ensures stability at long times. The resulting dynamics are characterized by an initial energy loading phase followed by a controlled relaxation into a protected subspace, rather than direct preparation of a stationary dark state.

In this sense, funnel states are not optimal for infinite-time storage, but they provide an efficient mechanism for accessing high-energy configurations whose decay is constrained by the dissipative structure of the system.

\section{Conclusions}

We have introduced a Davies-MS framework for the systematic identification of long-lived energy storage states in interacting multilevel quantum batteries. Applied to a minimal two-qutrit model, the method reveals a hierarchy of dissipative roles—dark, funnel, and bright states—that cannot be inferred from symmetry arguments alone.

In this framework, the multilevel ladder structure enables a specific class of excited states, termed ''funnel states,'' which decay exclusively into protected manifolds. While not optimal for infinite-time storage, these states provide a practical route to accessing higher-energy protected subspaces without requiring the direct preparation of dark states. The Davies-MS decomposition further reveals the dominant decay pathways governing energy loss, offering a principled basis for designing targeted control and protection strategies in future implementations.

The framework naturally generalizes to larger systems and higher-dimensional local Hilbert spaces, where the number of protected states increases rapidly. Our results suggest that optimal quantum battery operation is achieved not solely by preparing dark states, but by exploiting structured dissipative pathways that connect high-energy states to protected manifolds. This perspective opens a route toward scalable and robust quantum energy storage in superconducting circuits and other multilevel architectures.

\section*{Acknowledgements}
The authors are grateful to the anonymous referees for their constructive suggestions, which significantly improved the clarity and experimental grounding of the manuscript. The authors acknowledge the use of Gemini Pro for language polishing and for improving the clarity and organization of explanatory text. All scientific content, analytical derivations, numerical calculations, figures, interpretations, citations, and conclusions were checked, revised, and approved by the authors, who take full responsibility for the manuscript. Y.P. acknowledges the financial support from the Scientific and Technological Research Council of Türkiye (TÜBİTAK) under the 2210-C Priority Areas National MSc Scholarship Program.
\appendix
\section{Basis transformation and the $N=2$ manifold}
To clarify, we present the Hamiltonian $\mathbf H$ of the $N=2$ subspace in two ways in Section ~\ref{section_3_b}. Equations~\ref{eq:H2bare} and~\ref{eq:H2SAbasis} are not different physical limits or an emergent physical degeneracy; they are related by a simple unitary change of basis. We also denote as $\mathbf H^{(2)}_{\text{bare}}$ the matrix representation of the two-qutrit Hamiltonian in the bare multi-atom basis $\{|20\rangle, |11\rangle, |02\rangle\}$. We introduce the symmetry-adapted basis
\(\{|A_2\rangle,|S_2\rangle,|11\rangle\}\), where
\(|A_2\rangle=(|20\rangle-|02\rangle)/\sqrt{2}\) and
\(|S_2\rangle=(|20\rangle+|02\rangle)/\sqrt{2}\). Therefore, the unitary transformation $U$ which transforms the bare states to the symmetry-adapted states $\{|A2\rangle$, $|S2\rangle$, $|11\rangle\}$ is given by:
\begin{equation}
    U = \begin{pmatrix} 1/\sqrt{2} & 1/\sqrt{2} & 0 \\ 0 & 0 & 1 \\ -1/\sqrt{2} & 1/\sqrt{2} & 0 \end{pmatrix}
\end{equation}
The transformed Hamiltonian is known as the symmetry-adapted Hamiltonian $\mathbf H^{(2)}_{\text{sym}}$, and is given by $\mathbf H^{(2)}_{\text{sym}}=U^{\dagger} \mathbf H_{\text{bare}}^{(2)}U$. By definition, bare and symmetry-adapted Hamiltonians have the same physical spectra (the same eigenvalues). This basis invariance preserves all the spectral characteristics and makes the actual invariant subspace (the $|A2\rangle$ component) explicit.

\section{Sensitivity to population redistribution within the degenerate sector}

\begin{figure}
    \centering
     \includegraphics[width=0.95\linewidth]{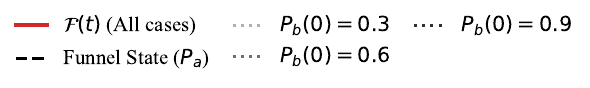}
     \includegraphics[width=0.95\linewidth]{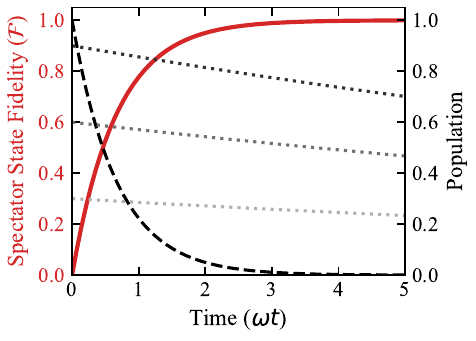}
    \caption{Spectator-state fidelity as a function of time for different initial population assignments in an orthogonal sector. The overlap of the long-time fidelity plateaus shows that the target-state fidelity is governed by the population transferred through the funnel channel and is insensitive to the specific population distribution within the orthogonal subspace. The parameters used in this simulation are dimensionless and chosen generically to illustrate the asymptotic insensitivity of the target state fidelity to the initial population distribution.}
    \label{fig:fidelity_population}
\end{figure}
We numerically tested the spectator state fidelity against controlled redistributions of the initial population in the degenerate subspace to directly address the physical impact of the spectral degeneracy in the $N=2$ manifold. We define the initial state density matrix $\rho(0)$ such that the total population in the symmetric sector is fixed, but redistributed among the orthogonal degenerate states. The time evolution of the target-state fidelity $\mathcal{F}(t)$ for different initial population weights is shown in Fig.~\ref{fig:fidelity_population}. The steady-state plateaus overlap perfectly showing that the fidelity is a function of only the total population transferred through the funnel channel into the protected state and is intrinsically insensitive to the specific coordinate distribution within the orthogonal degenerate subspace.
\section{Analytical checks of the dynamics}
\subsection{Dark-state and funnel conditions:}
The structure of the collective dissipator analytically guarantees the robustness of the steady-state plateaus observed in our numerical results. A state is perfectly robust to dissipation for a system coupled to a common bath with collective jump operator $L = (a_A + a_B)$ if it is in the kernel of $L$. A case in point is the $N=1$ antisymmetric target state $|D_1\rangle = \frac{1}{\sqrt{2}}(|10\rangle - |01\rangle)$, which has:
\begin{equation}
    L|D_1\rangle = \frac{1}{\sqrt{2}} (a_A + a_B)(|10\rangle - |01\rangle) = \frac{1}{\sqrt{2}} (|00\rangle - |00\rangle) = 0.
\end{equation}
 This confirms that the state is an exact eigenstate of the dissipative block. Another case is the funnel state we mentioned in Eq.~\ref{eq:funnel_condition}. Acting the collective dissipator $L$ on this funnel state on the $N=2$ manifold yields
\begin{equation}
L|A_2\rangle=\frac{1}{\sqrt2}(a_A+a_B)(|20\rangle-|02\rangle)=|10\rangle-|01\rangle=\sqrt2|D_1\rangle
\end{equation}
demonstrating the anti-symmetric state in the $N=2$ manifold behaves as a funnel state that funnels into the lower manifold dark state $|D_1\rangle$ under the collective dissipator.
\subsection{Short-time expansion of stored energy:}
The rapid initial energy uptake at $t=0$ warrants a closer look. To trace its physical origin, we examine the short-time expansion of the stored energy under the total Hamiltonian, $H = H_S + H_{\text{charge}}$. The relevant expectation value is straightforward: $E(t) = \text{Tr}(H_S \rho(t))$. Assuming initialization in the ground state $\rho(0) = |G\rangle\langle G|$, the early density matrix evolution follows the standard expansion:$$\rho(t) \approx \rho(0) - i t [H, \rho(0)] - \frac{t^2}{2} [H, [H, \rho(0)]]$$Notice the first-order term. It vanishes completely. Because the coherent drive operator $H_{\text{charge}}$ is entirely off-diagonal within the system's eigenbasis, it forces $\text{Tr}(H_S [H_{\text{charge}}, \rho(0)]) = 0$. The lowest-order non-zero contribution must therefore come from the second derivative. Evaluating the trace of that double commutator leaves a strictly positive result, one that scales directly with the squared effective driving amplitude:$$\Delta E(t) \approx \frac{t^2}{2} \text{Tr}\left( H_S [H_{\text{charge}}, [\rho(0), H_{\text{charge}}]] \right) \propto (\Omega_{\text{eff}} t)^2$$This $t^2$ dependence tells a clear physical story. At the very onset of the charging process, the stored energy undergoes coherent parabolic growth. The system essentially outpaces dissipation, bypassing any initial lag before environmental decay channels can fully set in.
\subsection{Perturbative scaling in the weak-anharmonicity limit:}
In the two-excitation sector, the ideal MS spectator state is broken by local anharmonicity $\alpha$ to an exact eigenstate. We consider the local anharmonicity term $H_{\text{anh}}$ (whose energy spectrum is defined in Eq.~\ref{eq:En}) as a perturbation to the exchange interaction $J$. The first order correction of the spectator state $|A_2\rangle$ comes from the admixture from the symmetric sector $|\phi_k\rangle:$
\begin{equation}
|\psi_{\text{exact}}\rangle \approx |A_2\rangle + \sum_k \frac{\langle \phi_k | H_{\text{anh}} | A_2\rangle}{\Delta E_k} |\phi_k\rangle.
\end{equation}
We can see that the energy denominator $\Delta E_k$ scales with the exchange coupling $J$, so the amplitude of the admixture scales as $\alpha/J$. Therefore, the loss in fidelity, which goes as the square of the overlap, has a perturbative reduction of order $(\alpha/J)^2$ for $\alpha/J \ll 1$.

\subsection*{4. Limiting cases relevant to Figs.~2 and 3}

\paragraph*{Harmonic limit, $\alpha=0$.}
In the harmonic limit, the two-qutrit Hamiltonian has an enhanced
ladder symmetry. The collective dark and spectator combinations obtained
from the MS decomposition are aligned with eigenspaces of $H_S$.
Consequently, after the charging field is switched off, the protected
component evolves only by an overall phase and remains in the kernel of
the collective dissipator. This explains the nearly stationary
long-time plateaus observed for the low-anharmonicity curves in
Figs.~\ref{fig:two_qutrit} and ~\ref{fig:qubit_qutrit}.

\paragraph*{Single-excitation target in Fig.~3.}
For the $N=1$ target state
\[
\ket{D_1}=\frac{1}{\sqrt{2}}(\ket{10}-\ket{01}),
\]
the local anharmonicity term proportional to $n(n-1)$ vanishes
identically because only occupations $n=0,1$ occur. Therefore
\[
H_S\ket{D_1}=(\omega-J)\ket{D_1},\qquad L\ket{D_1}=0.
\]
Thus $\ket{D_1}$ is simultaneously an eigenstate of $H_S$ and a dark
state of the collective dissipator for all $\alpha$. This gives an
analytical explanation of the flat post-cutoff fidelity plateau in
Fig.~\ref{fig:qubit_qutrit_b}.

\paragraph*{Two-excitation reference state in Fig.~2.}
The $N=2$ spectator reference state contains the components
$\ket{20}$, $\ket{11}$, and $\ket{02}$. For $\alpha\neq0$, the
anharmonic term shifts these components unequally, so the ideal MS
spectator combination is generally not an exact eigenstate of $H_S$.
After the drive is removed, its projection therefore contains phase
factors $e^{-i(E_m-E_n)t}$ between dressed components, producing
coherent beating in the fidelity. The corresponding stored energy,
however, is evaluated as $\Delta E(t)=\mathrm{Tr}[H_S\rho(t)]-E_{\rm gs}$.
Its unitary part is conserved after cutoff, and the remaining decay is
due to the dissipator acting on nonprotected population.
\paragraph*{Strong-anharmonicity limit, $\alpha/J\gg1$.}
For large anharmonicity, the upper transmon transition is detuned and
population transfer into the two-excitation manifold is suppressed.
Equivalently, the mixing between the ideal MS spectator/funnel sector and
the exact Hamiltonian eigenstates is no longer perturbatively small.
The qutrit then behaves increasingly like an effective two-level system,
which explains the reduction of the multilevel storage advantage in the
large-$\alpha$ curves in Fig~\ref{fig:two_qutrit_a}.

\bibliography{mybib} 
\end{document}